\documentclass[aps,pra,twocolumn,groupedaddress]{revtex4}

\usepackage{graphicx}
\usepackage{amsmath}
\usepackage{bm}
\usepackage{color}

\newcommand{\eq}[1]{(\ref{#1})}
\newcommand{\Eq}[1]{Eq.~(\ref{#1})}
\newcommand{\Figure}[1]{Figure~\ref{#1}}
\newcommand{\Fig}[1]{Fig.~\ref{#1}}
\newcommand{\Sec}[1]{Sec.~\ref{#1}}
\newcommand{\Ref}[1]{Ref.~\cite{#1}}
\def\beq{\begin{equation}} \def\eeq{\end{equation}}
\def\bea{\begin{eqnarray}} \def\eea{\end{eqnarray}}
\def\bse{\begin{subequations}} \def\ese{\end{subequations}}

   \def\vecr{{\bm r}} \def\vecv{{\bm v}}\def\vecD{{\bm \nabla}}
    
\def\||{\parallel}

\def\<{\left\langle} \def\>{\right\rangle}
\def\({\left(} \def\){\right)}

\begin{document}


\title{Domain-area distribution anomaly in segregating multicomponent superfluids}


\author{Hiromitsu Takeuchi}
\email{hirotake@sci.osaka-cu.ac.jp}
\homepage{http://hiromitsu-takeuchi.appspot.com}
\affiliation{Department of Physics, Osaka City University, 3-3-138 Sugimoto, Sumiyoshi-ku, Osaka 558-8585, Japan}


\date{\today}

\begin{abstract}
The domain-area distribution in the phase transition dynamics of ${\rm Z}_2$ symmetry breaking
is studied theoretically and numerically for segregating binary Bose--Einstein condensates in quasi-two-dimensional systems.
 Due to the dynamic scaling law of the phase ordering kinetics,
 the domain-area distribution is described by a universal function of the domain area, rescaled by the mean distance between domain walls. 
The scaling theory for general coarsening dynamics in two dimensions hypothesizes that
the distribution during the coarsening dynamics has a hierarchy with the two scaling regimes,
 the microscopic and macroscopic regimes with distinct power-law exponents.
The power law in the macroscopic regime, where the domain size is larger than the mean distance, is universally represented with the Fisher's exponent of the percolation theory in two dimensions.
On the other hand, the power-law exponent in the microscopic regime is sensitive to the microscopic dynamics of the system.
This conjecture is confirmed by large-scale numerical simulations of the coupled Gross--Pitaevskii equation for binary condensates.
In the numerical experiments of the superfluid system,
 the exponent in the microscopic regime anomalously reaches to its theoretical upper limit of the general scaling theory.
The anomaly comes from the quantum-fluid effect in the presence of circular vortex sheets, described by the hydrodynamic approximation neglecting the fluid compressibility.
It is also found that the distribution of superfluid circulation along vortex sheets obeys a dynamic scaling law with different power-law exponents in the two regimes.
An analogy to quantum turbulence on the hierarchy of vorticity distribution and the applicability to chiral superfluid $^3$He in a slab are also discussed.
\end{abstract}


\maketitle

\section{INTRODUCTION\label{Intro}}
In the phase transition dynamics of spontaneous symmetry breaking (SSB),
a number of topological defects are nucleated, forming a complicated network or texture in order parameter fields. 
This type of phenomenon can occur universally in systems ranging from condensed matter to cosmology and high energy physics \cite{1995Vilenkin, 2000Bunkov, 2002Onuki, 2006Vachaspati}.
The dynamic-scaling law in phase ordering kinetics hypothesizes that the growth of order parameter fields preserves the statistical similarity of the spatial patterns during such a SSB development \cite{1994Bray}.
This law has been accepted empirically by observing that the structure factors or correlation functions of the fields {\it collapse onto} a universal function after rescaling length by the mean distance $l$ between topological defects, which obeys a power law $l \propto t^{1/z}$ with the dynamic exponent $z$.

Considering the long research history of the phase ordering kinetics,
 it is recent that the application of percolation theory \cite{1994Stauffer} to SSB development attracted attention \cite{2007Arenzon, 2008Sicilia, 2012Olejarz}, and such an application has become an important problem in statistical physics \cite{2016Cugliandolo}.
Since percolation theory reveals different statistical aspects of the phase ordering kinetics,
such studies could lead to a greater understanding of the highly non-equilibrium physics
 of SSB development.
For example, consider two-dimensional coarsening dynamics of ${\rm Z}_2$ symmetry breaking, which are the most fundamental problem of phase ordering kinetics \cite{1994Bray}.
In this system, the order parameter is a real scalar field and the topological defect is a linear object called a domain wall.
One significant prediction by the application of percolation theory is that the time development of the number distribution $\rho(S,t)$ of domains of area $S$ obeys a universal power law $\rho \propto S^{-\tau_F}$ with the Fisher exponent $\tau_F=187/91 \approx 2$ of two-dimensional percolation \cite{1994Stauffer}.

There has been a growing interest in the phase ordering kinetics of atomic Bose-Einstein condensates (BECs),
and the dynamic-scaling law in superfluid systems has been investigated theoretically in different situations \cite{1996Damle, 2007Mukerjee, 2012Takeuchi, 2013Kudo, 2013Karl, 2014Hofmann, 2015Kudo, 2016Williamson, 2017Karl}.
Recently, percolation theory has been applied to the segregation of binary BECs in quasi-two-dimensions, and the dynamic finite-size-scaling analysis revealed that the domain structures preserve the percolation criticality during ${\rm Z}_2$ symmetry breaking with the percolation threshold $p_c=0.5$ \cite{2015Takeuchi}.
Recently, the domain-area distribution in segregating superfluids was shown to obey Fisher's power law for large-scale structures \cite{2016Takeuchi, 2016Bourges}.
Interestingly, the distribution of domains smaller than the characteristic domain size $l$ in the superfluid systems shows anomalous behavior different from that observed in conventional coarsening systems of nonconserved and conserved fields \cite{2007Arenzon,2009Sicilia}.
These results suggest that the dynamic-scaling law holds even when the quantum-fluid effect in superfluid systems becomes important
since the small-scale structure strongly depends on the ``microscopic'' dynamics of the domain wall in the system under consideration.
This anomalous behavior of smaller domains indicates the existence of a different dynamic-scaling regime that reflects the quantum fluid dynamics, namely, ``microscopic'' dynamics in quantum fluids.
Can such a scaling regime firmly coexist with the universal scaling regime of percolation theory in superfluid coarsening systems?
If so, the critical exponent and the effect of the microscopic nature on the scaling behavior should be important for a deeper understanding of the physics of SSB development, {\it e.g.}, seeking different scaling relations.
These aspects were unclear in the previous works \cite{2016Takeuchi, 2016Bourges}.

\begin{figure}
\begin{center}
\includegraphics[width=1.0 \linewidth,keepaspectratio]{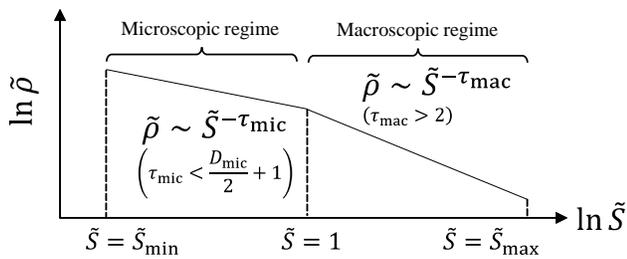}
\end{center}
\caption{(Color online)
The expected asymptotic form of the rescaled domain-area distribution $\tilde{\rho}$.
} 
\label{Fig_LogLog}
\end{figure}

To answer these questions,
we demonstrate the dynamic-scaling analysis according to the combined theory of phase ordering kinetics and percolation, called {\it phase ordering percolation}.
The theoretical conjecture presented here is summarized in \Fig{Fig_LogLog}.
Assuming an asymptotic form of the domain-area distribution,
this theoretical analysis is universally applicable to different coarsening systems.
The domain-area distribution has two distinct scaling regimes, namely, the microscopic and macroscopic regimes,
which are characterized by the exponents $\tau_{\rm mic}$ and $\tau_{\rm mac}=\tau_F$, respectively.
 Numerical experiments of binary BECs reveal that the microscopic exponent $\tau_{\rm mic}$ of the superfluid system
 reaches to the theoretical upper limit $3/2$,
 which is quite different from those of conventional coarsening systems.
This anomalous behavior is connected to the quantum-fluid effect of a circular vortex sheet by finding a dynamic scaling law derived from the quantum fluid dynamics.
Finally, an analogy to the hierarchy in quantum turbulence and the applicability of this scaling theory to the system of chiral-domain formation in quasi-two-dimensional superfluid $^3$He-A will be discussed.

\section{General scaling formalism}
We first formulate a generalized theory for evaluating the scaling behavior of phase ordering percolation
before discussing superfluid systems.

\subsection{Dynamic scaling of domain-area distribution}
 Consider the time evolution of domain-area distribution $\rho$ in a coarsening system that undergoes ${\rm Z}_2$ symmetry breaking in two dimensions.
There are two kinds of domains, namely, $\uparrow$- and $\downarrow$-domains.
We write the number of $\uparrow$- or $\downarrow$-domains, which have areas between $S$ and $S+dS$, divided by the system area $L^2$ at time $t$ as $\rho(S,t)dS$.

According to the dynamic scaling law in the phase ordering kinetics, domain structures of different times can be statistically similar by rescaling length with the mean inter-defect distance $l(t)$. By applying this empirical law to the domain-area distribution,
 $\rho(S,t)$ is described by a dimensionless universal function $\tilde{\rho}(\tilde{S})$ of $\tilde{S}=S/S_l$ with $S_l=\pi l^2$:
\begin{eqnarray}\label{rhoS}
\tilde{\rho}(\tilde{S}) = S_l^2\rho(S,t).
\end{eqnarray}
This is another expression of the dynamic-scaling law in the sense that 
the law is conventionally examined by observing the structure factor or the correlation function \cite{1994Bray}.
The dynamic-scaling law (\ref{rhoS}) has been experimentally \cite{2008Sicilia} and numerically \cite{2007Arenzon, 2009Sicilia, 2016Takeuchi, 2016Bourges} confirmed in different coarsening systems. 

\subsection{Normalization condition of wall length}
To analytically evaluate the rescaled distribution function (\ref{rhoS}),
it is useful to introduce the normalization conditions or sum rules for the domain-area distribution.
The condition for the total domain area was demonstrated in Ref. \cite{2007Arenzon}.
Here, we formulate the condition for the total domain-wall length 
$$
R(t)=L^2/l(t)
$$
 that yields useful input for the analysis of microscopic exponent $\tau_{\rm mic}$.

Because domain walls exist between $\uparrow$- and $\downarrow$-domains, 
the length $R$ is calculated by integrating the total length of the walls surrounding all $\uparrow$- or $\downarrow$-domains.
Thus, by introducing the length $l_{\rm w}(S)$ of the wall surrounding a domain of area $S$,
we have the normalization condition
\begin{eqnarray}\label{normalL}
\frac{R}{L^2}=\int_{S_{\rm min}}^{S_{\rm max}}  l_{\rm w} \rho dS = \frac{1}{\pi l}\int_{\tilde{S}_{\rm min}}^{\tilde{S}_{\rm max}}  \tilde{l}_{\rm w} \tilde{\rho} d\tilde{S}
\end{eqnarray}
with $\tilde{l}_{\rm w}\equiv l_{\rm w}/l$ and
\begin{eqnarray}\label{S_min_S_max}
\tilde{S}_{\rm min}\equiv \frac{S_{\rm min}}{S_l} \sim \(\frac{l_{\rm min}}{l}\)^2,~~~
\tilde{S}_{\rm max}\equiv \frac{S_{\rm max}}{S_l} \sim \(\frac{L}{l}\)^{\frac{91}{48}}.
\end{eqnarray}
 The lower cutoff $S_{\rm min}$ is determined by the microscopic length of the system, that is, by the thickness $l_{\rm min}$ of the domain wall; a domain is ill-defined if its area is smaller than
$$
S_{\rm min}=\pi l_{\rm min}^2.
$$
The upper cutoff $S_{\rm max}$ comes from percolation theory, applied after rescaling $L$ by $l$ with the effective system size $\tilde{L}=L/l$ \cite{2015Takeuchi}.
Here, I mention briefly this point although $S_{\rm max}$ is not so important to the main topic of this work. 
According to the percolation theory in two dimensions,
the largest domain is called the percolating cluster or domain with a non-trivial fractal dimension.
The area $S_{\rm max}$ of the percolating domain is connected with the percolation probability $P(p)$ at percolation threshold $p=p_c$, $P(p=p_c)=\lim_{L \to \infty} \frac{S_{\rm max}}{L^2}=\lim_{L \to \infty} \frac{S_{\rm max}/S_l}{L^2/S_l}\sim \tilde{L}^{-\beta/\nu}$, with the exponents $\beta=5/36$ and $\nu=4/3$ for two-dimensional percolation \cite{1994Stauffer}.
Here, $p_c=0.5$ is assumed for the coarsening dynamics of conventional systems \cite{2007Arenzon,2008Sicilia,2012Olejarz, 2016Cugliandolo}, which has been numerically confirmed for segregating binary BECs \cite{2015Takeuchi}.

\subsection{Microscopic and macroscopic regimes}
We assume an asymptotic form for $\tilde{S}_{\rm min} \ll S \ll 1$ and $1 \ll S \ll \tilde{S}_{\rm max}$ (see \Fig{Fig_LogLog}),
\begin{eqnarray}\label{tau_D}
 \tilde{\rho} \sim \tilde{S}^{-\tau},~~~\tilde{l}_{\rm w} \sim \tilde{S}^{D/2} 
\end{eqnarray}
with different exponents $\tau$ and $D$ in the two scaling regimes:
\begin{eqnarray}\label{tau_D_micmac}
(\tau, D)=\left\{
\begin{array}{ccc}
(\tau_{\rm mic}, D_{\rm mic}) & {\rm for}& \tilde{S}_{\rm min} \ll \tilde{S} \ll 1  \\
(\tau_{\rm mac}, D_{\rm mac}) &  {\rm for}& 1 \ll \tilde{S} \ll \tilde{S}_{\rm max}
\end{array} 
\right.
\end{eqnarray}
The exponent $D$ characterizes the fractal behavior of the domain walls, and thus
\begin{eqnarray}\label{Eq_D}
1 \leq D \leq 2.
\end{eqnarray}
The trivial exponent, $D=1$, is realized for circular domains with a relation
\begin{eqnarray}\label{Eq_Dtrivial}
l_{\rm w} = 2\sqrt{\pi S} \Rightarrow \tilde{l}_{\rm w} = 2\pi \sqrt{ \tilde{S}} \Rightarrow  \tilde{l}_{\rm w} \sim  \tilde{S}^{0.5}.
\end{eqnarray}
The upper bound, $D=2$, comes from the spatial dimension of our system.

Under the above assumption, multiplying \Eq{normalL} by $l$, one obtains
\begin{eqnarray}\label{normalL_reduce}
\lambda_{\rm mic}+\lambda_{\rm mac}+\lambda_{\rm P} \sim 1
\end{eqnarray}
with 
$$
\lambda_{\rm mic}=\int^{1}_{\tilde{S}_{\rm min}}  \tilde{l}_{\rm w}\tilde{\rho} d\tilde{S},~~~\lambda_{\rm mac}=\int_{1}^{\tilde{S}_{\rm max}} \tilde{l}_{\rm w}\tilde{\rho} d\tilde{S}.
$$
Here, the contribution from the domain walls surrounding the largest (percolating) domain is extracted explicitly as 
$$
\lambda_P=l_{\rm w}(S_{\max})/R
$$
 in \Eq{normalL_reduce}.

\subsection{Restrictions on dynamic scaling exponents}
The theoretical restrictions for $\tau_{\rm mic}$ and $\tau_{\rm mac}$ are obtained by evaluating the condition (\ref{normalL_reduce}) in the limit of $\tilde{S}_{\rm min} \to 0$ and $\tilde{S}_{\rm max} \to \infty$.
All terms on the left-hand side of \Eq{normalL_reduce} must be on the order of unity or zero.
From the condition \eq{Eq_D}, $\tilde{l}_{\rm w}(S_{\rm max})$ is smaller than or on the order of $\tilde{S}_{\rm max}$, and thus $\tilde{l}_{\rm w}(S_{\rm max})  \lesssim \tilde{S}_{\rm max}\sim (L/l)^{91/48}$.
Then we have $\lambda_P =  l \tilde{l}_{\rm w}(S_{\rm max}) /R \lesssim l (L/l)^{91/48}/(L^2/l)=(l/L)^{2-91/48} \to 0$ in the limit.
 Substituting \Eq{tau_D} into $\lambda_{\rm mic}$, one obtains 
$\lambda_{\rm mic} \sim \int^{1}_{\tilde{S}_{\rm min}}  \tilde{S}^{D_{\rm mic}/2-\tau_{\rm mic}} d\tilde{S}$, and thus
we have the restriction for $\tau_{\rm mic}$ as
\begin{eqnarray}\label{Rest_mic}
\tau_{\rm mic} < D_{\rm mic}/2+1.
\end{eqnarray}
Similarly, the restriction for $\tau_{\rm mac}$ is derived as $\tau_{\rm mac} > D_{\rm mac}/2+1$.
A similar evaluation is available for the area condition $1/2=\int S\rho dS$,
where the integral represents the total area $L^2/2$ occupied by $\uparrow$- or $\downarrow$-domains, divided by the system area $L^2$.
This condition yields $\tau_{\rm mic} < 2$ and
\begin{eqnarray}\label{Rest_mac}
\tau_{\rm mac} > 2.
\end{eqnarray}
The restriction (\ref{Rest_mac}) has been obtained in \Ref{2007Arenzon} and is consistent with the prediction of the percolation theory: $\tau_{\rm mac}=\tau_F=187/91 >2$.

To validate our theory, the obtained restriction (\ref{Rest_mic}) in the microscopic regime is compared with results in conventional coarsening systems of nonconserved and conserved order parameters at zero temperature \cite{2007Arenzon, 2009Sicilia}.
For nonconserved coarsening in the two-dimensional Ising model (2dIM) \cite{2007Arenzon}, 
the distribution function $\rho$ becomes flat for $S \to 0$ with $\tau_{\rm mic}=0$.
On the other hand, 2dIM simulations for the conserved case \cite{2009Sicilia} indicate $\tau_{\rm mic}=-0.5$.
Smaller domains are close to or indeed circular with $D_{\rm mic} = 1.1$ and $D_{\rm mic} = 1$ for the former and latter cases, respectively;
 the restriction (\ref{Rest_mic}) is safely satisfied.
These results suggest that $\tau_{\rm mic}$ varies depending on the ``microscopic'' dynamics of the domain wall in different types of coarsening systems, 
 while $\tau_{\rm mac}$ takes the universal value $\tau_{\rm mac}=\tau_F$ for conventional systems \cite{2007Arenzon, 2009Sicilia} as was mentioned in the introduction.

\section{Numerical experiments of segregating binary superfluids}\label{Num}
\begin{figure*}
\begin{center}
\includegraphics[width=1. \linewidth,keepaspectratio]{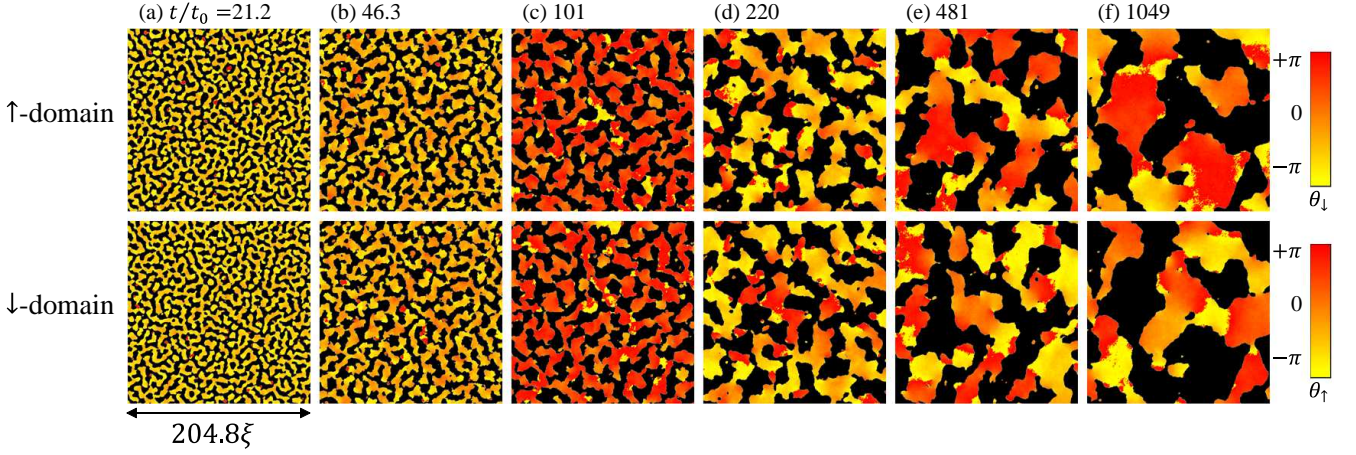}
\end{center}
\caption{(Color online)
A time evolution of domain coarsening dynamics in a numerical experiment of segregating binary BECs.
Black regions show $\uparrow$-domains ($\downarrow$-domains) from $t/t_0=$21.2 to $t/t_0=$1049 in upper (lower) panels.
 The gradation represents spatial distribution of the phase $\theta_{\downarrow(\uparrow)}={\rm arg}\Psi_{\downarrow(\uparrow)}$ in $\downarrow(\uparrow)$-domains.
 A vortex is represented by a end point of branch cut (jump from $\theta_{\uparrow,\downarrow}=-\pi$ to $\pi$).
 The phase $\theta_{\uparrow(\downarrow)}$ is nearly homogeneous in some of small $\uparrow(\downarrow)$-domains ($S <S_l$), at which branch cuts of $\theta_{\downarrow(\uparrow)}$ end, by forming vortex sheets along their domain walls.
} 
\label{Fig_Develop}
\end{figure*}

Our main goal is to study the realizability of the microscopic regime in superfluid systems together with the universal macroscopic regime and identify the scaling behavior.
Fortunately, the numerical experiment of segregating binary BECs \cite{2015Takeuchi,2016Takeuchi} can be used to achieve this goal.
Figure \ref{Fig_Develop} shows a time evolution of $\uparrow$- and $\downarrow$-domains in a numerical experiment of segregating binary BECs.

\subsection{Segregation of binary condensates}
The segregation dynamics were simulated by solving the coupled Gross--Pitaevskii (GP) equations \cite{2008Pethick} derived from the GP Lagrangian in a quasi-two-dimensional system,
\begin{eqnarray}\label{Eq_Lagrangian}
{\cal L}=\int d^2x \( {\cal P}_{\uparrow}(\Psi_\uparrow)+{\cal P}_{\downarrow}(\Psi_\downarrow)-g_{\uparrow\downarrow}n_{\uparrow} n_{\downarrow} \).
\end{eqnarray}
Here, we used
$$
{\cal P}_j=-\hbar n_j \partial_t \theta_j-\frac{\hbar^2}{2m}(\vecD \sqrt{n_j})^2-\frac{m}{2}n_j\vecv_j^2-\frac{g_{jj}}{2}n_j^2
$$
 with atomic mass $m$, complex order parameter $\Psi_j=\sqrt{n_j}e^{i\theta_j}$, and superfluid velocity $\vecv_j=\frac{\hbar}{m}\vecD \theta_j$ of the $j$-component  $(j=\uparrow, \downarrow)$.
The ${\rm Z}_2$ symmetry breaking is caused by the dynamic instability starting from the initial state with a constant density  $|\Psi_{\uparrow}|^2=|\Psi_{\downarrow}|^2=n/2={\rm const}$ and zero velocity $\vecv_j =0$ for the coupling constants $g_{\uparrow\downarrow}>g_{\uparrow\uparrow}=g_{\downarrow\downarrow}=g>0$.
For strong segregation in our simulation with $g_{\uparrow\downarrow}/g=2$,
the domain-wall thickness  is on the order of the healing length
\begin{eqnarray}\label{Eq_xi}
\xi \equiv \frac{\hbar}{\sqrt{gmn}},
\end{eqnarray}
 and we set $l_{\rm min}= \xi$. 
For making the segregation dynamics of binary BECs,
 the coupling constants are controlled by changing the oblateness of the trapped condensates, using techniques of magnetic Feshbach resonance \cite{2008Papp, 2010Tojo} and dressed states \cite{2011Lin, 2015Nicklas}, and changing the internal states of atoms \cite{2016Eto}.
 Since spinor BECs can be described effectively by the same hydrodynamics,
the system is also available for examining the scaling behaviors, demonstrated here, in cold atom systems.
 See also the references in {\it e.g.}, \Ref{2016Williamson}.

The dynamic instability, triggered by adding small random seeds to the initial state, develops into complicated domain patterns with characteristic length $l$ (see \Fig{Fig_Develop}).
We investigated the domain-area distribution after the time ($t=\tau_0 \sim t_0\equiv \frac{\hbar}{gn}$) when domain patterns emerge clearly in the initial stage ($0< t \lesssim \tau_0$) of the SSB development.
The characteristic length $l_0$ of the initial domain patterns at that time is on the order of $\xi$ as is determined from the Bogoliubov theory \cite{2015Takeuchi}.
For weakly segregated systems, $l_0$ and $l_{\rm min}$ diverge as $\propto 1/\sqrt{g_{\uparrow\downarrow}/g-1}$.
Then, the characteristic time $\tau_0$, at which the initial domain pattern emerges, becomes larger as $\tau_0 \sim\frac{t_0}{g_{\uparrow\downarrow}/g-1} \ln \frac{n}{2\delta^2}$.
Here, $\delta^2$ is the amplitude of the density fluctuation due to the initial random seeds made by a white noise or Gaussian one.  
To realize the scaling behaviors for a finite-size system of binary BECs with a finite life time,
it is better to use the systems of stronger inter-component interaction with smaller $l_0$, $\tau_0$, and $l_{\rm min}$.
In this way, we require a computational system whose size is much larger to satisfy the conditions $\tilde{S}_{\rm min} \ll 1$ and $\tilde{S}_{\rm max} \gg 1$ for attaining the coexistence of the microscopic and macroscopic regimes.
In our simulation, the size and number of the numerical grids are set to meet these requirements during the time development.
For more details on numerical analysis, see Appendix A.

\subsection{Dynamic scaling plot of domain-area distribution}
\begin{figure}
\begin{center}
\includegraphics[width=1. \linewidth,keepaspectratio]{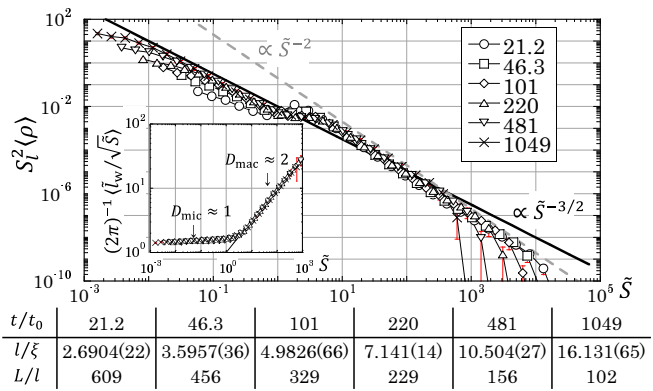}
\end{center}
\caption{(Color online)
Dynamic-scaling plot of the averaged domain-area distribution $\langle \rho (S) \rangle$.
The graph legends represent the time from $t/t_0=$21.1 to 1049 with $t_0=\frac{\hbar}{gn}$.
The table provides information on the corresponding values of $l/\xi$ and $L/l$.
The inset shows the corresponding scaling plot for $\tilde{l}_{\rm w}$ from $t/t_0=$ 101 to 1049 with the solid line $\tilde{l}_{\rm w}=2\pi\tilde{S}$.
Error bars correspond to the standard deviation $S_l^2\sqrt{\delta_X^2}$ $\left[(2\pi)^{-1}\sqrt{\delta_X^2}\right]$ of the ensemble average $S_l^2\langle X \rangle$ $\left[ (2\pi)^{-1}\langle X \rangle \right]$ for the quantity $X$ $\left( =\rho~{\rm or}~\tilde{l}_{\rm w}/\sqrt{\tilde{S}} \right)$,
where the average $\langle X \rangle$ and its variance $\delta_X^2=\langle \(X-\langle X \rangle\)^2\rangle$ is calculated over the 64 samples of the numerical simulations.
} 
\label{Fig_s_rho}
\end{figure}

\begin{figure}
\begin{center}
\includegraphics[width=1.0 \linewidth,keepaspectratio]{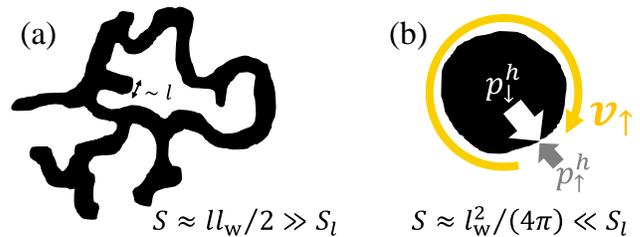}
\end{center}
\caption{(Color online)
Schematic of typical shapes of (a) a larger domain ($S\gg S_l$) and (b) a smaller domain ($S \ll S_l$) that appear in a domain structure in the SSB development.
The black curves shows domain walls whose thickness $l_{\rm min}$ is much smaller than the mean distance $l$ between domain walls in the late stage of the SSB development.
(a) A domain in the macroscopic regime is like a snaky and branching trail of width $\sim l$ and length $\sim l_{\rm w}/2$; the area of a trail is given by $S\approx ll_{\rm w}/2$.
This relation between $S$ and $l_{\rm w}$ is consistent with the dynamic scaling relation $\tilde{l}_{\rm w}=2\pi \tilde{S} \Rightarrow l_{\rm w}/l=2\pi S/S_l \Rightarrow S=l l_{\rm w}/2$ of the macroscopic regime in the inset of \Fig{Fig_s_rho}.  
(b) The domain is almost circular by obeying the relation of \Eq{Eq_Dtrivial} approximately; $S\approx l_{\rm w}^2/(4\pi)$.
The $\uparrow$-domain (white region) contains a vortex with a circulation $\kappa n_{\rm v}$,
 whose core is occupied by a circular $\downarrow$-domain (black region) at rest.
The hydrostatic pressure of the $\uparrow$-domain on the wall is smaller than that of the $\downarrow$-domain, $p_\uparrow^h< p_\downarrow^h$,  in the presence of the circular superflow ($|\vecv_\uparrow|=\frac{\kappa |n_{\rm v}|}{2\pi R}$). 
} 
\label{Fig_CircleDomain}
\end{figure}

\Figure{Fig_s_rho} shows rescaled plots of the time evolution of the domain-area distribution.
Because we have $l\sim l_0$ ($\sim \xi$), that is, $S_l \sim S_{\rm min}$ in the initial domain patterns,
 there are fewer domains in the microscopic regime,
where the size of domains is comparable to or smaller than the domain-wall thickness $l_{\rm min}\sim \xi$ and then domains are ill-defined physically (see also the discussion on a core-less vortex in \Sec{Sec:QFE}).
Thus, in the early stage  ($t/\tau_0\lesssim t/t_0 \lesssim 220$ with $l/\xi \sim 1$), the power-law behavior (\ref{tau_D}) of the microscopic regime is ill-established and the rescaled plots of different times do not coincide for $\tilde{S} \lesssim 1$,
although the scaling behavior appears clearly in the macroscopic regime with $\tau_{\rm mac} \approx \tau_F$.
In the late stage ($t/t_0 \gtrsim 220$ with $l/\xi \gtrsim 10$), a power-law behavior becomes well-established in the microscopic regime too with $\tau_{\rm mic}\approx 3/2$.
 Then plots of different times {\it collapse onto} a universal curve of the asymptotic form of \Fig{Fig_LogLog} in both regimes.
 The dynamic-scaling plot for $\tilde{l}_{\rm w}$ becomes successful also in the late stage (see \Fig{Fig_s_rho} inset).
We have exponents $D_{\rm mic}\approx 1$ and $D_{\rm mac}\approx 2$, which is similar to the results of conventional systems \cite{2009Sicilia}.
\Figure{Fig_CircleDomain} illustrates schematically the typical domain shapes in the macroscopic and microscopic regimes.

The exponent $\tau_{\rm mic}$ reaches to the upper limit, $3/2$ of the restriction (\ref{Rest_mic}) with $D_{\rm mic}=1$, that is,
the number of domains is maximized in the superfluid system.
This anomalous behavior in the microscopic regime is in contrast to the behavior of the above-mentioned conventional systems,
while the macroscopic regime shows the universal behavior of the percolation criticality.

The configuration of the domain-area distribution in the connection regime ($\tilde{S}\sim 1$)
 can differ depending on coarsening systems and the initial conditions.
We see a fine structure around the connection regime in \Fig{Fig_s_rho}.
However, the restrictions (\ref{Rest_mic}) and (\ref{Rest_mac}) are obtained independently of the connection regime.
Thus, the detailed structure of the connection regime is not relevant to our discussion on the scaling exponents $\tau$ and $D$ in the limits of $\tilde{S}_{\rm min} \to 0$ and $\tilde{S}_{\rm max} \to \infty$.

\section{A circular vortex sheet as a quantum-fluid effect\label{Sec:QFE}}
Generally speaking, the dynamic scaling behavior in the microscopic regime of the domain-area distribution reflects the ``microscopic'' dynamics of domain walls in the system under consideration.
An effective theory that describes the dynamics in our system is quantum fluid dynamics for multi-component superfluids.
It is interesting how the quantum-fluid effect is connected to the anomalous behavior of the microscopic regime in the quantum fluids.
Here, the anomaly is suggested to occur in the presence of circular vortex sheets described in the quantum fluid dynamics.

\subsection{Hydrodynamic description}\label{Sec_Hy_Des}

To provide a quantitative evidence of the above suggestion,
we introduce a hydrodynamic theory for a vortex sheet between two domains in a two-component superfluid.
Consider a core-less vortex of circulation $\kappa n_{\rm v}~(n_{\rm v}=0,\pm 1,\pm 2,...)$ with the circulation quantum $\kappa$ in the superfluid,
 where the core of a vortex in a sufficiently large $\uparrow$-domain is occupied by a $\downarrow$-domain at rest as is illustrated in \Fig{Fig_CircleDomain}~(b).
 The domain may be not completely circular in the highly non-equilibrium SSB development
 while the stationary solution of a core-less vortex has a circular domain wall.
 Since the $\uparrow$-domain has a circular superflow along a domain wall between $\uparrow$- and $\downarrow$-domains, 
a relative rotational velocity between the domains causes a distribution of vorticity along the wall forming a circular vortex sheet.
For example, see Fig.~2 in \Ref{2013Hayashi}.

Here, we shall describe a circular vortex sheet as an equilibrium state.
According to a theory of quantum fluid dynamics \cite{2013Hayashi},
such a state is stabilized by the centrifugal force caused by the rotational superflow,
while a flat vortex sheet is dynamically unstable without external forces owing to the Kelvin--Helmholtz instability (KHI) \cite{2010Takeuchi}.
The radius $R$ of the circular wall is estimated by the equation of pressure equilibrium \cite{Landau_statphys}
\begin{eqnarray}\label{Epre}
\Delta p^h=p^h_{\downarrow}-p^h_{\uparrow}=\frac{\sigma_{\rm wall}}{R}
\end{eqnarray}
 with the tension coefficient $\sigma_{\rm wall}$ of the wall and the hydrostatic pressure $p^h_j$ ($j=\uparrow,\downarrow$) along the wall in the $j$-domain.
In quantum fluid dynamics for the GP Lagrangian \eq{Eq_Lagrangian}, there is a so-called quantum pressure originating from the uncertainty relation or the spatial gradient of the order parameter amplitude. The quantum pressure is included in the tension coefficient $\sigma_{\rm wall}$ and  neglected in the pressure $p^h_j$
for our hydrodynamic theory, where the fluid compressibility is neglected in bulk.

 The pressure $\bar{p}^h_j$ in the bulk, where the superfluid is at rest, is the sum of $p^h_j$ and the hydrodynamic pressure $p^d_j=\frac{1}{2}\rho_jv_j^2$; $\bar{p}^h_j=p^h_j+p^d_j$ according to the Bernoulli's principle.
 Here, $\rho_j$ and $v_j$ are the mass density and superfluid velocity along the domain wall in the $j$-domain, respectively.
 The ${\rm Z}_2$ symmetry of the Hamiltonian for the multi-component superfluid corresponds to $\bar{p}^h_\uparrow=\bar{p}^h_\downarrow$,
which reduces to 
\begin{eqnarray}\label{E_Hpre}
\Delta p^h=p^d_\uparrow -p^d_\downarrow=\frac{1}{2}\rho_\uparrow v_\uparrow^2-\frac{1}{2}\rho_\downarrow v_\downarrow^2.
\end{eqnarray}
For example, in the case of segregated binary BECs,
the hydrostatic pressure is given by the Lagrangian density ${\cal P}_j$ of the $j$-component.
In the incompressible approximation of the hydrodynamic theory neglecting the quantum pressure term $\propto (\nabla \sqrt{n_j})^2$,
 one obtains ${\cal P}_j \to p_j^h= \mu_j n_j-\frac{1}{2}\rho_jv_j^2-\frac{1}{2}gn_j^2=\bar{p}^h_j-p^d_j$ with $\bar{p}^h_j=\mu_j n_j-\frac{1}{2}gn_j^2$ for a stationary state $\Psi_j(\vecr,t)=e^{-i\mu_j t/\hbar}\Phi_j(\vecr)$,
 where $\mu_j$ is the chemical potential of the $j$-component.
 Because of the original ${\rm Z}_2$-symmetry of the Lagrangian,
we have
\begin{eqnarray}\label{Eq_p_hs}
\bar{p}^h_\uparrow=\bar{p}^h_\downarrow=\frac{1}{2}gn^2
\end{eqnarray}
 with $\mu_j=gn$,
with which we have the two ground states with the same energy; $(|\Psi_\uparrow|^2, |\Psi_\downarrow|^2)=(n,0)$ or $(0,n)$ in bulk, corresponding to spontaneous breaking of ${\rm Z}_2$-symmetry.
Then, we obtain \Eq{E_Hpre} for the case of binary BECs too.
Note that the relation \eq{E_Hpre} is generally derived from the hydrodynamic theory under the condition $\bar{p}^h_\uparrow=\bar{p}^h_\downarrow$ of the ${\rm Z}_2$-symmetry.

\subsection{Dynamic scaling of a circular vortex sheet}\label{DSCVS}

The hydrodynamic theory gives the dynamic-scaling relation between the circulation number $n_{\rm v}$ and the area $S=\pi R^2$ of a circular domain in the microscopic regime.
 A dynamic scaling law for superfluid circulation $\kappa n_{\rm v}$ is obtained from Eqs.~\eq{Epre} and \eq{E_Hpre}.
 Since the $\downarrow$-domain is at rest ($v_{\downarrow}^2=0$) and the $\uparrow$-domain has a superfluid velocity $v_{\uparrow}^2=(\kappa n_{\rm v}/2\pi R)^2$ along the circular domain wall,
 one obtains $n_{\rm v}^2=2(2\pi/\kappa)^2\sigma_{\rm wall}R/ \rho_{\uparrow}$, 
which reduces to the dynamic scaling law for the vortex distribution,
\begin{eqnarray}\label{n_to_S}
 | \tilde{n}_{\rm v}(\tilde{S}) | \equiv \sqrt{\frac{l_{\rm min}}{l}}|n_{\rm v}(S)| \sim \tilde{S}^{1/4}~~~(\tilde{S}_{\rm min} \ll \tilde{S}\ll 1).
\end{eqnarray}
 Here, we used the fact that the length
\begin{eqnarray}\label{Eq_xiwall} 
\xi_{\rm wall} \equiv \frac{1}{2}\( \frac{\kappa}{2\pi}\)^2 \frac{\rho_{\uparrow}}{\sigma_{\rm wall}}
\end{eqnarray}
 characterizes a length scale related to the domain wall, and thus, it should be on the order of the wall thickness $l_{\rm min}$.

In the conventional coarsening systems, a circular domain ($D_{\rm mic}=1$) shrinks and finally collapses in bulk owing to dissipation or evaporation.
In contrast, a circular domain with vortices ($n_{\rm v} \neq 0$) in the quantum fluid is stable with a rotational superflow in equilibrium.
This is why the number of domains in the microscopic regime is statistically enhanced up to the upper limit $\tau_{\rm mic} \to 3/2$ in the superfluid system.

To apply the dynamic scaling law \eq{n_to_S} without any change,
we assume that circular domains in the microscopic regime are {\it independent} as was assumed in the literatures \cite{2007Arenzon, 2009Sicilia}.
More specifically, the domains in the regime are not expected to have holes of opposite component within.
In other words, there are no domains embedded inside a circular domain in the microscopic regime.
Actually, we could not find such small domains embedded inside domains with $S \ll S_l$ in the numerical experiments [see {\it e.g.}, \Fig{Fig_Develop}(e,f)].

\subsection{Dynamic scaling plot of circulation number}\label{Sec_DSP}

\begin{figure}
\begin{center}
\includegraphics[width=1.0 \linewidth,keepaspectratio]{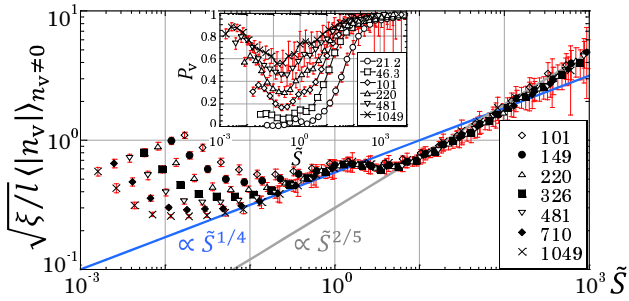}
\end{center}
\caption{(Color online)
Dynamic scaling plot of the circulation quantum number $|n_{\rm v}(S)|$ from $t/t_0=$101 to 1049.
The inset shows the ratio $P_{\rm v}(\tilde{S})$ of the number of domains with $|n_{\rm v}|\neq 0$ from $t/t_0=$ 21.2 to 1049.
Error bars correspond to the standard deviation for the quantities $\sqrt{\xi/l} |n_{\rm v}|$ and $P_{\rm v}$.}
\label{Fig_s_nv}
\end{figure}

We examined the dynamic scaling law (\ref{n_to_S}) for the numerical experiments of binary BECs. 
\Figure{Fig_s_nv}  shows the dynamic scaling plot of the ensemble average of $|n_{\rm v}(S)|$ for domains with $n_{\rm v}(S)\neq 0$.
The number $n_{\rm v}$ was obtained by numerically integrating the superfluid velocity along domain walls.
 The scaling law (\ref{n_to_S}) is observed in the late stage for the same reason as is the case of the scaling plot of \Fig{Fig_s_rho}.
It is interesting that the dynamic scaling plot is successful even in the macroscopic regime with $|\tilde{n}_{\rm v}|\approx \tilde{S}^{2/5}$.

 Each plot of different times in \Fig{Fig_s_nv} has an upturn at $\tilde{S} \sim \tilde{S}_{\rm min}$.
 The upturn effect reflects the core structure of a core-less vortex whose core has a radius on the order of $l_{\rm min}$, which was neglected in the process of the derivation of \Eq{n_to_S}.
 The correction for the upturn effect is computed qualitatively by considering the density variation owing to the rotational superflow in the stationary solution of a core-less vortex for the GP Lagrangian \eq{Eq_Lagrangian}.
 Considering the situation of \Fig{Fig_CircleDomain}~(b) for $S \gtrsim S_{\rm min}$ and neglecting the quantum pressure term $\propto \( \vecD \sqrt{n_j} \)^2$ again,
we may write as $gn_{\uparrow}=\mu_\uparrow-\frac{\hbar^2 n_{\rm v}^2}{2m  r^2}$ ($r>R \gtrsim \xi$) and $gn_{\downarrow}=\mu_\downarrow$ ($r < R$) in the stationary vortex solution.
Then, $\Delta p^h=\frac{1}{2}gn_{\downarrow}^2-\frac{1}{2}gn_{\uparrow}^2=\frac{1}{2}gn^2\left[1-\(1- \frac{n_{\rm v}^2\xi^2}{2R^2} \)^2 \right]=\frac{\sigma_{\rm wall}}{R}$
 reduces to
\begin{eqnarray}\label{Eq_nvGP}
|n_{\rm v}| = \frac{R}{\xi} \sqrt{2\( 1-\sqrt{1-\frac{\xi^2}{\xi_{\rm wall} R}} \)}
\end{eqnarray}
This formula consistently reproduces the upturn effect around $R \sim l_{\rm min}$ ($S \sim S_{\rm min}$) with $l_{\rm min}=\xi\sim \xi_{\rm wall}$.
Moreover, \Eq{Eq_nvGP} reduces to the dynamic scaling law \eq{n_to_S} for $l_{\rm min} \ll R$ ($S_{\rm min} \ll S$) approximately.

\subsection{Vortex supply by Kelvin--Helmholtz instability\label{KHI}}
The above statistical discussion is meaningful when domains with vortices are dominant over vortex-free domains ($n_{\rm v} = 0$) in the microscopic regime.
The inset of \Fig{Fig_s_nv} shows the plot of the probability $P_{\rm v}(\tilde{S})$ in which a domain of $\tilde{S}$ has vortices with $n_{\rm v} \neq 0$.
The probability is very small in the initial stage and then grows in the early stage.
In the late stage, most domains contain vortices.

One might assume that quantized vortices are nucleated due to the Kibble--Zurek mechanism \cite{1976Kibble,1985Zurek}.
The mechanism causes nucleation of quantized vortices via a spatially inhomogeneous growth of a complex scalar field of the superfluid order parameter from zero in the initial state.
 In our system, however, the superfluid order parameters $\Psi_\uparrow$ and $\Psi_\downarrow$ are finite initially.
Instead, a real scalar field, {\it e.g.}, $n_\uparrow-n_\downarrow$, can be our effective order parameter,
 where not vortices but domain walls are nucleated as topological defects of this system.
In fact, the phase $\theta_{\uparrow,\downarrow}$ is almost uniform and there are less vortices in earlier times [see \Fig{Fig_Develop}(a)].

How are quantized vortices that exist in the late stage nucleated?
A possible mechanism for the vortex nucleation is the KHI.
The KHI can occur when there is a vortex sheet between two fluids.
If the relative velocity across the vortex sheet exceeds the critical velocity $V_{\rm KH}$ for the KHI in binary BECs,
a ripple wave is excited on the domain wall and the vortex sheet releases a portion of `vorticity charge' as quantized vortices (see e.g., Fig.~1 in \Ref{2010Takeuchi} for the dynamics under external potentials).
A vortex sheet can exist due to a local superflow induced by a complex motions of domain walls during the highly non-equilibrium development.
Since the critical velocity $V_{\rm KH}$ is zero without external potential,
 quantized vortices are easily nucleated from vortex sheet in our system.
 Note that $V_{\rm KH}$ is nonzero for the circular vortex sheet in the microscopic regime
 since the centrifugal force due to the rotational superflow plays the role of the stabilizing force.

In the initial stage, relative motion between two components is negligible
since the Bogoliubov excitations, amplified due to the dynamic instability from the fully mixed state ($\Psi_\uparrow= \Psi_\downarrow={\rm const}$),
 does not cause momentum exchange between two components \cite{2011Ishino}.
This is why there are less vortices nucleated by the KHI in earlier times [\Fig{Fig_Develop}(a,b)]
 but we found vortices in later times [\Fig{Fig_Develop}(e,f)] after domain walls start to make a complex motion by inducing a relative velocity locally.

The released vortices due to the KHI can be absorbed again by domain walls.
When vortices are absorbed into a domain in the microscopic regime by forming a vortex sheet,
the sheet can be stabilized  against the KHI as its equilibrium condition was described by Eq.~(\ref{Epre}).
Such a circular vortex sheet survives long unless the domain collides to other domains.
 As a result, the probability $P_{\rm v}$ grows with time, becoming substantially large in the late stage,
 where almost domains contain vortices.
This result clearly shows that the influence of superflow is essential to understand the later-stage dynamics.

\section{Summary and discussion\label{Sec_Sum}}
A dynamic scaling hypothesis for the domain-area distribution during coarsening dynamics of ${\rm Z}_2$ symmetry breaking was proposed (\Fig{Fig_LogLog}).
The large-scale numerical simulations of segregating binary superfluids supported the prediction of the hypothesis for the late stage of the SSB development:
 there exist the two scaling regimes with distinct exponents $\tau_{\rm mic}\approx 3/2$ and $\tau_{\rm mac}\approx 2$ in the microscopic and macroscopic regimes, respectively (\Fig{Fig_s_rho}).
 Such a hierarchy was a missing piece of the puzzle numerically and theoretically in the previous studies with smaller systems \cite{2016Takeuchi, 2016Bourges}.
The number of domains in the microscopic regime is anomalously maximized in the superfluid systems 
while the macroscopic regime exhibits the universal scaling behavior of percolation theory.
The dynamic scaling law \eq{n_to_S} for the vortex distribution and its dynamic scaling plot (\Fig{Fig_s_nv}) showed that the anomaly is induced by the quantum-fluid effect in the presence of circular vortex sheets in the microscopic regime.
Since a large-scale system of binary or spinor BECs has been realized for investigating non-equilibrium fluid dynamics in quasi-two-dimensions \cite{2017Kim},
it is in great hope that the dynamic scaling behaviors of segregating binary superfluids will be observed in future experiments,
which would pioneer the research on a different type of non-equilibrium statistical mechanics raised in the quantum systems.

Interestingly, the dynamic-scaling plot for circulation number $n_{\rm v}$ yields a different scaling behavior with $|\tilde{n}_{\rm v}| \sim \tilde{S}^{2/5}$ in the macroscopic regime.
This suggests that distribution of velocity or vorticity may obey a different scaling law or power law for length scales larger than the average distance $l$ between domain walls.
This situation is similar to the hierarchy in quantum turbulence,
 where the Kolmogorov or semi-classical power law is realized over length scales larger than the average distance between vortex lines and a different power law is expected for smaller length scales \cite{2014Barenghi, 2017Walmsley}.
In the field of quantum turbulence, the connection regime between the two scaling regimes has also received a lot of attention and should be discussed further qualitatively here.
An excitation of Kelvin wave, a perturbation mode of a quantized vortex, due to reconnections of vortices is crucial in the context of the transition from the Kolmogorov cascade in the `macroscopic' (semi-classical) regime to the Kelvin wave  cascade in the `microscopic' (quantum) regime \cite{2007Lvov, 2008Kozik}.
By using the above analogy between quantum turbulence and our system,
an excitation of ripple wave, a perturbation mode of a domain wall, due to collisions of domain walls can be important to describe the connection regime.
This situation is in contrast to the conventional coarsening systems of nonconserved field where collisions of domain walls occur rarely and most of domain walls are smooth curves owing to the energy dissipation.  
A hierarchy in turbulent superflow has never been observed experimentally in a direct way,
 and then the scaling behavior for smaller length scale have never been identified. 
In this sense, theoretical and experimental investigations into these aspects are fruitful for the domain coarsening system of binary BECs.

To reinforce our hypothesis on phase ordering dynamics,
the theory can be applied to different multi-component superfluids,
where vortex sheets are stabilized.
An important application is quasi-two-dimensional $^3$He-A confined in a slab system,
where chiral domain walls were recently visualized \cite{2016Sasaki}. 
The order parameter is the vector field $\hat{\bm l}$ that represents the direction of orbital angular momentum of the Cooper pair in the dipole-locked $^3$He-A.
A one-to-one correspondence exists between the $\hat{\bm l}$ field of $^3$He-A and the (pseudo-)spin field of spinor (binary) BECs, according to the Mermin--Ho relation that represents vorticity distribution due to the vector-field texture \cite{2003Volovik,2005Kasamatsu,2012Volovik}.
In this sense, these superfluid systems can show a similar effect even if the ``microscopic'' dynamics differs between the $^3$He-A and spinor (binary) BECs.
By quenching the $^3$He-A system to the superfluid phase from the normal fluid phase,
domain growth should occur.
Then, the characteristic domain-area distribution, as is illustrated in \Fig{Fig_LogLog}, will be observed when the characteristic domain size is much larger than thickness of domain wall and much smaller than the system size.

\begin{acknowledgments}
We are grateful to L. Cugliandolo, L. A. Williamson, P. B. Blakie, O. Ishikawa, K. Kasamatsu, A.~Oguri, and S.~Inouye  for useful discussion and comments on this work.
This work was supported by JSPS KAKENHI Grant Numbers JP26870500, JP17K05549, JP17H02938.
The present research was also supported in part by the Osaka City University (OCU) Strategic Research Grant 2017  for young researchers.
\end{acknowledgments}



\appendix

\section{Technical description on the numerical analysis\label{App_Num}}
The numerical simulation of the coupled Gross-Pitaevskii (GP) equations is computed on a two-dimensional lattice of square grids.
The grid size $\Delta x$ must be smaller than $\xi$ to precisely simulate the dynamics of the instability and the quantized vortices, which will appear at the late stage.
We require a computational system whose size is much larger to satisfy the conditions $\tilde{S}_{\rm min} \ll 1$ and $\tilde{S}_{\rm max} \gg 1$ for attaining the coexistence of the microscopic and macroscopic regimes.
In our simulation, which utilized the Crank--Nicolson method with periodic boundary conditions,
 the size and number of the numerical grids are set to meet these requirements during the time development as $\Delta x/\xi=0.4$ and $L/\Delta x=4096$, respectively.

The numerical results are obtained by averaging $64$ samples of the time evolution.
The ensemble averages, except for the average of $l$, were taken by considering the periodic boundary conditions; averaged quantities are calculated by averaging 64 samples after the secondary average over the $8\times 8$ pseudo-samples obtained by shifting the field data of a single simulation in the $x$ and $y$ directions by $i \times L/8$ and $j \times L/8$ with integers $i$ and $j$ ($0\leq i,j \leq 7$), respectively. The length $L/8$ is much larger than the characteristic length $l$ of the field, so the secondary average improves the statistical analysis substantially.

The domain-area distribution was calculated as follows.
A domain wall is defined as a collection of sides between neighboring grids with $n_{\rm d}\equiv n_\uparrow-n_\downarrow > 0$ and $n_{\rm d} < 0$.
A domain is surrounded by a closed domain wall (or the system boundary and a open domain wall that ends at the boundary).
A saddle point, an intersection of the walls in the numerical lattice,
occasionally occurs when two domain walls are close to each other.
Then, we calculate the average value $\bar{n}_{\rm d}$ of $n_{\rm d}$ over the four grids around the
saddle point and then regard the point with $\bar{n}_{\rm d} > 0$ ($<0$) as a
point occupied by the $\uparrow$($\downarrow$)-component connecting the
diagonal domains with $n_{\rm d} > 0$ ($<0$).
All domains are labeled with different numbers.
The area of each domain is calculated by counting the number of grids existing inside the domain.
By performing a histogram analysis for the number distribution of domain area,
we obtain the domain-area distribution after averaging the data over the samples.





\begin{thebibliography}{99}
\bibitem{1995Vilenkin} A. Vilenkin and E. P. S. Shellard, {\it Cosmic Strings and Other Topological Defects} (Cambridge University Press, Cambridge, 1995).
\bibitem{2000Bunkov} {\it Topological Defects and the Non-Equilibrium Dynamics of Symmetry Breaking Phase Transitions}, edited by Y. M. Bunkov and H. Godfrin, Vol. 549 of NATO Advance Science Institute, Series C: Mathematical and Physical Sciences (Kluwer, Dordrecht, 2000).
\bibitem{2002Onuki} A. Onuki, {\it Phase Transition Dynamics} (Cambridge University Press, Cambridge, 2002).
\bibitem{2006Vachaspati} T. Vachaspati, {\it Kinks and Domain Walls: An Introduction to Classical and Quantum Solitons} (Cambridge University Press, Cambridge, 2006).
\bibitem{1994Bray} A. J. Bray, Adv. Phys. {\bf 43}, 357 (1994).



\bibitem{1994Stauffer} D. Stauffer and A. Aharony, {\it An Introduction to Percolation}, 2nd ed. (Taylor and Francis, London, 1994). 
\bibitem{2007Arenzon} J. J. Arenzon, A. J. Bray, L. F. Cugliandolo, and A. Sicilia, Phys. Rev. Lett. {\bf 98}, 145701 (2007); A. Sicilia, J. J. Arenzon, A. J. Bray, and L. F. Cugliandolo, Phys. Rev. E {\bf 76}, 061116 (2007).
\bibitem{2008Sicilia} A. Sicilia, J. J. Arenzon, I. Dierking, A. J. Bray, L. F. Cugliandolo, J. Mart\'inez-Perdiguero, I. Alonso, and I. C. Pintre, Phys. Rev. Lett. {\bf 101}, 197801 (2008).

\bibitem{2012Olejarz} J. Olejarz, P. L. Krapivsky, and S. Redner, Phys. Rev. Lett. {\bf 109}, 195702 (2012).
\bibitem{2016Cugliandolo} A. Tartaglia, L. F. Cugliandolo, and M. Picco, Europhys. Lett. {\bf 1116}, 26001 (2016); 
L. Cugliandolo, {\it Phase ordering kinetics, aggregation and percolation in two dimensions}, Plenary lecture in STATPHYS26, Lyon, France (2016).



\bibitem{1996Damle} K. Damle, S. N. Majumdar, and S. Sachdev, Phys. Rev. A {\bf 54}, 5037 (1996).
\bibitem{2007Mukerjee} S. Mukerjee, C. Xu, and J. E. Moore, Phys. Rev. B {\bf 76}, 104519 (2007).
\bibitem{2012Takeuchi} H. Takeuchi, K. Kasamatsu, M. Tsubota, and M. Nitta, Phys. Rev. Lett. {\bf 109}, 245301 (2012).
\bibitem{2013Kudo} K. Kudo and Y. Kawaguchi, Phys. Rev. A {\bf 88}, 013630 (2013).
\bibitem{2013Karl} M. Karl, B. Nowak, and T. Gasenzer, Phys. Rev. A {\bf 88}, 063615 (2013).
\bibitem{2015Kudo} K. Kudo and Y. Kawaguchi, Phys. Rev. A {\bf 91}, 053609 (2015).
\bibitem{2014Hofmann} J. Hofmann, S. S. Natu, and S. Das Sarma, Phys. Rev. Lett. {\bf 113}, 095702 (2014).
\bibitem{2016Williamson} L. A. Williamson and P. B. Blakie, Phys. Rev. Lett. {\bf 116}, 025301 (2016); Phys. Rev. A {\bf 94}, 023608 (2016).
\bibitem{2017Karl} Markus Karl and Thomas Gasenzer, New J. Phys. {\bf 19} 093014 (2017).
\bibitem{2015Takeuchi} H. Takeuchi, Y. Mizuno, and K. Dehara, Phys. Rev. A {\bf 92}, 043608 (2015). 
\bibitem{2016Takeuchi} H. Takeuchi, J. Low Temp. Phys. {\bf 183}, 169 (2016). 
\bibitem{2016Bourges} A. Bourges and P. B. Blakie, Phys. Rev. A {\bf 95}, 023616 (2017).
\bibitem{2009Sicilia} A. Sicilia, Y. Sarrazin, J. J. Arenzon, A. J. Bray, and L. F. Cugliandolo, Phys. Rev. E {\bf 80}, 031121 (2009).


\bibitem{2008Pethick} C. J. Pethick and H. Smith, {\it Bose--Einstein Condensation in Dilute Gases}, 2nd ed. (Cambridge University Press, Cambridge, 2008).

\bibitem{2008Papp} S. B. Papp, J. M. Pino, and C. E. Wieman, Phys. Rev. Lett. {\bf 101}, 040402 (2008).
\bibitem{2010Tojo} S. Tojo, Y. Taguchi, Y. Masuyama, T. Hayashi, H. Saito, and T. Hirano,  Phys. Rev. A {\bf 82} 033609 (2010).
\bibitem{2011Lin} Y.-J. Lin, K. Jim\'enez-Garc\'ia, and I. B. Spielman, Nature {\bf 471} 83 (2011).
\bibitem{2015Nicklas} E. Nicklas, M. Karl, M. H\"ofer, A. Johnson, W. Muessel, H. Strobel, J. Tomkovi\u{c}, T. Gasenzer, and M. K. Oberthaler, Phys. Rev. Lett. {\bf 115}, 245301 (2015).
\bibitem{2016Eto} Yujiro Eto, Masahiro Takahashi, Masaya Kunimi, Hiroki Saito, and Takuya Hirano, New J. Phys. {\bf 18}, 073029 (2016).

\bibitem{2013Hayashi} S. Hayashi, M. Tsubota, and H. Takeuchi, Phys. Rev. A, {\bf 87}, 063628 (2013).

\bibitem{2010Takeuchi} H. Takeuchi, N. Suzuki, K. Kasamatsu, H. Saito, and M. Tsubota,
Phys. Rev. B {\bf 81}, 094517 (2010).

\bibitem{Landau_statphys} L. D. Landau and E. M. Lifshitz, {\it Statistical Physics: Part I} (Pergamon, Oxford, 1980), Chap. 15.

\bibitem{1976Kibble} T. W. B. Kibble, J. Phys. A {\bf 9}, 1387 (1976).

\bibitem{1985Zurek} W. H. Zurek, Nature (London) {\bf 317}, 505 (1985); Phys. Rep. {\bf 276}, 177 (1996).

\bibitem{2011Ishino} S. Ishino, M. Tsubota, H. Takeuchi, Phys. Rev. A {\bf 83}, 063602 (2011).

\bibitem{2017Kim} Joon Hyun Kim, Sang Won Seo, Yong-il Shin, Phys. Rev. Lett. {\bf 119}, 185302 (2017). 


\bibitem{2014Barenghi} C. F. Barenghi, L. Skrbek, and K. R. Sreenivasan, Proc. Natl. Acad. Sci. USA {\bf 111}, 4647 (2014).
\bibitem{2017Walmsley}P. M. Walmsley and A. I. Golov, Phys. Rev. Lett. {\bf 118}, 134501 (2017).

\bibitem{2007Lvov} V. S. L'vov, S. V. Nazarenko, and O. Rudenko, Phys. Rev. B, {\bf 76}, 024520 (2007).
\bibitem{2008Kozik} E. Kozik and B. Svistunov, Phys. Rev. B, {\bf 77}, 060502 (2008).

\bibitem{2016Sasaki} Y. Sasaki, {\it Visualizing textural domain walls in superfluid $^3$He by Magnetic Resonance Imaging}, Invited talk in the International Conference on Quantum Fluids and Solids 2016 (QFS2016), Prague, Czech Republic (2016).

\bibitem{2003Volovik} G. E. Volovik, {\it The Universe in a Helium Droplet} (Clarendon Press,
Oxford, 2003).
\bibitem{2005Kasamatsu} K. Kasamatsu, M. Tsubota, and M. Ueda, Phys. Rev. A {\bf 71}, 043611 (2005).
\bibitem{2012Volovik} G. E. Volovik and M. Krusius, Physics {\bf 5}, 130 (2012).


\end{thebibliography}
\end{document}